\documentclass[a4paper]{jpconf}
\usepackage{amssymb}
\usepackage{amsthm}
\usepackage{latexsym}
\newcommand{\be}{\begin{eqnarray}}
\newcommand{\ee}{\end{eqnarray}}
\newcommand{\bdm}{\begin{displaymath}}
\newcommand{\edm}{\end{displaymath}}

\begin{document}
\title{\textbf{The metric-affine formalism of  $f(R)$ gravity}}
\author{\textbf{Thomas P.~Sotiriou and Stefano Liberati}}
\date{}
\address{SISSA-International School of Advanced Studies, via Beirut 2-4, 34014, Trieste, Italy and INFN, Sezione di Trieste}
\ead{\textbf{sotiriou@sissa.it; liberati@sissa.it}}

%%%%%%%%%%%%%%%%%%%%%%%%%%%%%%%%%%%%%%%%%%%%%%%%%%%%%%%
\begin{abstract}
Recently a class of alternative theories of gravity which goes under the name
f(R) gravity, has received considerable attention, mainly due to its
interesting applications in cosmology. However, the phenomenology of such
theories is not only relevant to cosmological scales,  especially when it is
treated within the framework of the so called Palatini variation, an
independent variation with respect to the metric and the connection, which is
not considered a priori to be the Levi-Civita connection of the metric. If this
connection has its standard geometrical meaning the resulting theory will be a
metric-affine theory of gravity, as will be discussed in this talk. The general
formalism will be presented and several aspects of the theory will be covered,
mainly focusing on the enriched phenomenology that such theories exhibit with
respect to General Relativity, relevant not only to large scales (cosmology)
but also to small scales (e.g. torsion).
\end{abstract}
%%%%%%%%%%%%%%%%%%%%%%%%%%%%%%%%%%%%%%%%%%%%%%%%%%%%%%%

\section{Introduction}
\label{intro}
%%%%%%%%%%%%%%%%%%%%%%%%%%%%%%%%%%%%%%%%%%%%%%%%%%%%%%%
It took a very long time before it was realized that Newtonian gravity was not the complete theory describing gravity but a limiting case of Einstein's theory, General Relativity (GR). This is not at all surprising since Newtonian gravity had successfully grasped all the basic aspects of the gravitational interaction ({\emph e.g.}~inverse square law) that appeared in current experiments. However, the passage from Newtonian gravity to GR was not triggered solely (some would say not even mainly) by the inability of the theory to explain some experimental data, but also for the theoretical motivation to present a theory that describes non-inertial frames, an extension of Special Relativity that can include gravity.

There are indication that GR might be facing similar difficulties with those Newton's theory faced almost 100 year's ago. We now have observations related to the late, low energy universe that indicate that it undergoes an accelerated expansion. In order to explain this picture with our standard theory of gravity we need to introduce some unobserved, exotic components in its energy budget, dark matter and dark energy. What is more these components sum up to approximately 96\% of the total energy of the universe, dark matter accounting for almost 20\% and dark energy for 76\% \cite{Tegmark:2006az}. If it was just for our inability to explain these observation we might not be intrigued to consider extending or modifying your theory of gravity. However, simple explanations within the framework of GR, such as the introduction of a cosmological constant to account for dark energy are burderned with serious problems, like the coincidence and the magnitude problems \cite{Carroll:2000fy}. At the same time these explanations are difficult to fit in our theoretical physics framework. Therefore, just as it happend with Newtonian gravity, motivations to modify GR come also from theoretical physics, and remarkably they were presented much before the observations discussed above were made ({\emph e.g.}~\cite{uti}).

 In particular, there are indications that the classical action of gravity admits higher order corrections with respect the linear term in the scalar curvature (Einstein--Hilbert action), either in the form of higher order curvature invariants or non minimally coupled scalar fields. This is the case if one takes into consideration quantum correction, or considers the classical gravitational action as an effective low energy action of some more fundamental theory like String/M-theory \cite{quant1,quant2,quant3,quant4,noji2,vassi}.

So, the question pending is whether by modifying our theory of gravity we can at the same time explain current observation without the need of unknown energy component and at the same time address the questions raised about gravity by theoretical physics. A noble but not very easy to achieve goal, as it has turned out. There are many tentatives towards this direction. Some examples are scalar-tensor theories \cite{faraoni}, the DGP model \cite{dvali}, brane-worlds \cite{marteens}, etc. Notice that all the above models modify GR by adding a new ingredient to it: an extra field or extra dimensions. 

Here we would like to follow a different path and present a theory of gravity in which no extra field or dimension is present and at the same time is as close as possible to the geometrical spirit of GR: metric-affine gravity.  In section \ref{action} we discuss the main features like the form of the action and its characteristic. We proceed in section \ref{fe} to derive the field equation. The coupling to matter and its implications (such as the presence of torsion and non-metricity leading to enriched phenomenology in all scales) is discussed. Section \ref{conc} contains conclusions.

%%%%%%%%%%%%%%%%%%%%%%%%%%%%%%%%%%%%%%%%%%%%%%%%%%%%%%%
\section{$f(R)$ action and projective invariance}
\label{action}
%%%%%%%%%%%%%%%%%%%%%%%%%%%%%%%%%%%%%%%%%%%%%%%%%%%%%%%
As already mentioned we are interested in a theory which is as close as possible to the geometrical spirit of GR, i.e. it describes a 4-manifold with a metric $g_{\mu\nu}$ which is used to measure distances and therefore defines the dot product of a vector $\xi$, $\|\xi\|^2=g_{\mu\nu}\xi^\mu\xi^\nu$, and a connection
$\Gamma^\lambda_{\phantom{a}\mu\nu}$ which defines parallel transport and therefore the covariant derivative, which we give here in index notation for clarity
\be 
\nabla_\mu A^\nu_{\phantom{a}\sigma}=\partial_\mu A^\nu_{\phantom{a}\sigma}+\Gamma^\nu_{\phantom{a}\alpha\mu} A^\alpha_{\phantom{a}\sigma}-\Gamma^\alpha_{\phantom{a}\mu\sigma} A^\nu_{\phantom{a}\alpha}.
\ee
$\partial_\mu$ denotes partial differentiation with respect to $x^\mu$. 

It is important to stress that it is the connection and not the metric one uses to define the Riemann tensor,
\be
\label{riemann}
R^\mu_{\phantom{a}\nu\sigma\lambda}=-\partial_\lambda\Gamma^\mu_{\phantom{a}\nu\sigma}+\partial_\sigma\Gamma^\mu_{\phantom{a}\nu\lambda}+\Gamma^\mu_{\phantom{a}\alpha\sigma}\Gamma^\alpha_{\phantom{a}\nu\lambda}-\Gamma^\mu_{\phantom{a}\alpha\lambda}\Gamma^\alpha_{\phantom{a}\nu\sigma},
\ee
and the Ricci tensor,
\be
\label{ricci}
R_{\mu\nu}=R^\lambda_{\phantom{a}\mu\lambda\nu}=\partial_\lambda \Gamma^\lambda_{\phantom{a}\mu\nu}-\partial_\nu \Gamma^\lambda_{\phantom{a}\mu\lambda}+\Gamma^\lambda_{\phantom{a}\sigma\lambda}\Gamma^\sigma_{\phantom{a}\mu\nu}-\Gamma^\lambda_{\phantom{a}\sigma\nu}\Gamma^{\sigma}_{\phantom{a}\mu\lambda}.
\ee 
In standard GR one imposes the covariant conservation of the metric, $\nabla_\lambda g_{\mu\nu}=0$, and the connection is then given with respect of the metric by the formula $\Gamma^\lambda_{\phantom{a}\mu\nu}=\frac{1}{2}g^{\lambda\sigma}(\partial_\mu g_{\nu\sigma}+\partial_\nu g_{\mu\sigma}-\partial_\sigma g_{\mu\nu})$, i.e. it is the Levi-Civita connection. Therefore, a contraction of the Ricci tensor given in eq.~(\ref{ricci}) with the metric leads to the usual Ricci scalar, which is a function of the metric, $R=R(g)$. Using this scalar we construct the Einstein-Hilbert action,
\be
S_{EH}=\frac{1}{2\kappa}\int d^4 x \sqrt{-g} R(g),
\ee
the variation of which with respect to the metric leads to the Einstein equations. $\kappa=8\pi G$ and $g$ denotes the determinant of the metric.

We want to abandon the assumption that the covariant derivative of the metric  vanishes, but keep the geometrical meaning of the connection as is, i.e. have the independent connection define parallel transport and the covariant derivative. The outcome will therefore be a metric-affine theory of gravity. Since, in this case the metric and the connection will be independent quantities we will have to vary the action with respect to both of them in order to derive the field equations \footnote{ This is called Palatini variation. However, metric-affine gravity should be confused with what is called Palatini formalism. Even though, in the Palatini formalism there exist an independent connection and the Palatini variation is used, this connection does not define parallel transport or the covariant derivatives \cite{sotlib,sot4}. Note also that metric-affine gravity has been studied from a different prospective, {\emph i.e.}~as a gauge theory of gravity (see for example \cite{rubi} for a study on $f(R)$ actions and \cite{hehlmccrea} for a thorough review).}. In GR we start from the Einstein--Hilbert action since we want an action that is built out of a generally covariant scalar and at the same time leads to second order field equations. $R(g)$ depends on the second derivative of the metric as well and would normally lead to fourth order field equation under standard metric variation, but remarkably higher order derivative gather up in a surface term \cite{landau}. This is not true for actions non-linear in $R(g)$. However, here we consider the connection and the metric as independent quantities and, therefore, $R$ does not contain second derivatives of the metric or the connections and any action of the form
\be
\label{maaction}
S=\frac{1}{2\kappa}\int d^4 x \sqrt{-g} f(R),
\ee
will lead to second order field equations. This is the action we will be using from now on. Note that an $f(R)$ action is a straightforward generalization of the Einstein--Hilbert one and is commonly used in modified gravity \cite{buh,staro,bar1,bar2,capo,capo2,carroll,noji,vollick,sot2,sot3,sot1} since, even though one cannot claim to get it directly as an exact low energy limit of a more fundamental theory, it seems to include many of the phenomenologically interesting terms appearing in such effective actions. Therefore, it can serve as a very good first approximation that we can use to get insight into the matter.  For a more detailed discussion on the properties of the action and a more detailed exploration of the possible form of an action suitable for metric-affine gravity see \cite{sotlib}.

Before going further, we need to notice a specific property of the action (\ref{maaction}) in metric-affine gravity. Recall that the connection is an independent quantity and we have supposed no symmetry. Let us consider the projective transformation
\be
\label{proj}
\Gamma^{\lambda}_{\phantom{a}\mu\nu}\rightarrow \Gamma^{\lambda}_{\phantom{a}\mu\nu}+{\delta^\lambda}_\mu\xi_\nu.
\ee
The Ricci tensor and scalar transform like
\be
\label{projRicci}
R_{\mu\nu}\rightarrow R_{\mu\nu}-2\partial_{[\mu}\xi_{\nu]}, \qquad R\rightarrow R,
\ee 
{\em i.e.}~the gravitational action (\ref{maaction}) is projectively invariant \cite{schro,sand,hehl}. However, the matter action is not projectively invariant, so if we do not find a way to break the invariance of the gravitational action we will not be able to derive consistent field equations.

As we see from eq.~(\ref{proj}) we could break this invariance by assuming that the connection is symmetric. However, we would like to allow the connection to have a non-symmetric part since as we will see later on this will allow the possibility of torsion. On other possibility would be to include extra terms in the action, such as $R^{\mu\nu}R_{\mu\nu}$,  which are not invariant under projective transformations. We will avoid this here, since first of all it goes beyond the realm of $f(R)$ gravity which we want to examine here, and secondly it is burdened with difficulties related to torsion analyzed in \cite{sotlib}. Therefore, in order to break the projective invariance we will have to use some constraint on the connection. Specifically we are going to set $S_{\mu}=S_{\sigma\mu}^{\phantom{ab}\sigma}=0$ where
\be
\Gamma^{\lambda}_{\phantom{a}[\mu\nu]}\equiv S_{\mu\nu}^{\phantom{ab}\lambda}\equiv\Gamma^{\lambda}_{\phantom{a}[\mu\nu]}
\ee
is the Cartan torsion tensor. The best way to impose this constraint is to add to the action the Lagrange multiplier
\be
\label{lm2}
S_{LM}=\int d^4 x \sqrt{-g} B^\mu S_{\mu},
\ee
as proposed for the Einstein-Hilbert action in \cite{sand} \footnote{See also \cite{sotlib} for a detailed study on alternatives in order to break the projective invariance of the gravitational action.}.

%%%%%%%%%%%%%%%%%%%%%%%%%%%%%%%%%%%%%%%%%%%%%%%%%%%%%%%
\section{Field equations and Matter fields}
\label{fe}
%%%%%%%%%%%%%%%%%%%%%%%%%%%%%%%%%%%%%%%%%%%%%%%%%%%%%%%

We will proceed to derive the field equations for metric-affine $f(R)$ gravity. Let us start by considering the variation of the matter action ${\cal S}_M$. In the usual way we will define the stress-energy tensor as 
\be
T_{\mu\nu}\equiv-\frac{2}{\sqrt{-g}}\frac{\partial {\cal S}_M}{\partial g^{\mu\nu}}.
\ee
Since the matter action is independently varied with respect to the connection as well we will also need here the quantity
\be
\Delta_{\lambda}^{\phantom{a}\mu\nu}\equiv-\frac{2}{\sqrt{-g}}\frac{\partial {\cal S}_M}{\partial \Gamma^\lambda_{\phantom{a}\mu\nu}},
\ee
which has been referred to in the past as ``hypermomentum'' \cite{hehl}.
%\be
%\delta S_M=\int d^4 x \frac{\partial S_M}{\partial g^{\mu\nu}}\delta g^{\mu\nu}+\int d^4 x \frac{\partial S_M}{\partial %\Gamma^\lambda_{\phantom{a}\mu\nu}}\delta \Gamma^\lambda_{\phantom{a}\mu\nu}
%\ee
%or in terms of $T_{\mu\nu}$ and $\Delta_{\lambda}^{\phantom{a}\mu\nu}$:
%\be
%\label{varmat}
%\delta_g S_M=-\frac{1}{2}\int d^4 x\sqrt{-g}T_{\mu\nu}\delta g^{\mu\nu}\quad
%\delta_\Gamma S_M=-\frac{1}{2}\int d^4 x\sqrt{-g}\Delta_{\lambda}^{\phantom{a}\mu\nu}\delta 
%\Gamma^\lambda_{\phantom{a}\mu\nu}
%\ee
If we denote $S_G$ as $S_G=S+S_{LM}$, add this quantity to the matter action and apply the stationary action principle we get
\be
0=\delta_g S_G+\delta_g S_M, \qquad\qquad
0=\delta_\Gamma S_G+\delta_\Gamma S_M,
\ee
and consequently after some mathematical manipulation the following field equations \cite{sotlib}:
\begin{eqnarray}
\label{field1t}
& &f'(R) R_{(\mu\nu)}-\frac{1}{2}f(R)g_{\mu\nu}=\kappa T_{\mu\nu}, \\
\label{field2t}
& &\frac{1}{\sqrt{-g}}\bigg[-\nabla_\lambda\left(\sqrt{-g}f'(R)g^{\mu\nu}\right)+\nabla_\sigma\left(\sqrt{-g}f'(R)g^{\mu\sigma}\right){\delta^\nu}_\lambda\bigg]+{}\nonumber\\ & &\qquad\qquad+2f'(R)g^{\mu\sigma}S^{\phantom{ab}\nu}_{\sigma\lambda}=\kappa(\Delta_{\lambda}^{\phantom{a}\mu\nu}-\frac{2}{3}\Delta_{\sigma}^{\phantom{a}\sigma[\nu}{\delta^{\mu]}}_{\lambda}),\\
\label{field3t}
& & S_{\mu\sigma}^{\phantom{ab}\sigma}=0.
\end{eqnarray}
Note that eq.~(\ref{field3t}) is the constraint we derive from the Lagrange multiplier (\ref{lm2}).

Having derived the field equations we can now examine the phenomenology of the theory by considering specific matter fields. Let us start with the electromagnetic field. Its action is
\be
S_{EM}=-\frac{1}{4}\int d^4 x \sqrt{-g} F^{\mu\nu}F_{\mu\nu},
\ee
leading to the usual stress-energy tensor
\be
T^{EM}_{\mu\nu}=F_\mu^{\phantom{a}\sigma}F_{\sigma\nu}-\frac{1}{4}g_{\mu\nu}F^{\alpha\beta}F_{\alpha\beta},
\ee
whereas $\Delta_{\lambda}^{\phantom{a}\mu\nu}=0$, since the matter action in independent of the connection \cite{sotlib}.
The field equations simplify dramatically in this case:
\begin{eqnarray} \label {elfe1}& & f'(R) R_{(\mu\nu)}-\frac{1}{2}f(R)g_{\mu\nu}=\kappa T_{\mu\nu}, \\ & & \label{elfe2}
\nabla_\lambda\left(\sqrt{-g}f'(R)g^{\mu\nu}\right)=0, \end{eqnarray}
and no room is left for torsion. Contracting eq.~(\ref{elfe1}) we get
\be \label{conelfe1} f'(R) R-2f(R)=0,\ee
which is an algebraic equation in $R$ with solutions $R=c_i$. We will not examine here cases in which  eq.~(\ref{conelfe1}) has no solutions or is identically satisfied, since such choices for $f(R)$ lead to problems in coupling matter to the theory \cite{ferr,sotlib}. Since now $R$ is a constant, so is $f(R)$ and $f'(R)$ and therefore eq.~(\ref{elfe2}) reduces to 
$\Gamma^\lambda_{\phantom{a}\mu\nu}=\{^\lambda_{\phantom{a}\mu\nu}\}$, i.e. the connection becomes the Levi-Civita one. Consequently $R_{(\mu\nu)}=R_{\mu\nu}(g_{\mu\nu})$ and we remain with a single field equation:
\be R_{\mu\nu}(g_{\mu\nu})-\frac{1}{4}c_i g_{\mu\nu}=\kappa' T^{EM}_{\mu\nu}, \quad \kappa'=\kappa/f'(c_i)
\ee
which is formally Einstein's equation. Needless to say that the whole procedure applies in vacuum as well.

We can also easily check what happens in the case of a perfect fluid. The stress energy tensor is as usual
\be
\label{pfse}
T^{\mu\nu}=(\rho+p)u^\mu u^\nu+p g^{\mu\nu},
\ee
where $p$ denotes the pressure, $u^\mu$ the 4-velocity and $\rho$ the energy density. $\Delta_{\lambda}^{\phantom{a}\mu\nu}=0$ once again since a perfect fluid can essentially be described in the level of the action by two scalar fields, its energy density and the velocity potential \cite{scha,stone,sotlib}. So, the field equations are identical to eqs.~(\ref{elfe1}) and (\ref{elfe2}) with $T_{\mu\nu}$ given by eq.~(\ref{pfse}). Contracting eq.~(\ref{elfe1}) we get
\be f'(R) R-2f(R)=\kappa T=-\kappa(\rho-3p),\ee
and we can easily distinguish the two standard cases:
\begin{itemize}
\item Dust: $p=0\quad \Rightarrow \quad f'(R) R-2f(R)=-\kappa\rho$
\item Radiation: $\rho=3p\quad \Rightarrow \quad f'(R) R-2f(R)=0\quad \Rightarrow \quad R=c_i$
\end{itemize}
Notice that in the case of radiation $R$ reduces to a constant as in the case of the electromagnetic field as expected. Recall that the relation between the metric and the connection is given by
\be \nabla_\lambda\left(\sqrt{-g}f'(R)g^{\mu\nu}\right)=0 \ee
which when $R$ is a constant reduces to the definition of the Levi-Civita connection. However, in the more general case, which includes dust, this equation reveals that there is non-metricity. 

We  just saw two cases of standard matter fields in which torsion vanishes. Additionally, a scalar field can fit into the description already given for a perfect fluid. One might also notice that in the case that $f(R)$ is linear the theory described above reduces to standard GR for these matter fields. But this should not mislead us to think that there is no significant difference from GR even when $f(R)=R$. Let us recall the full form of the second field equation for an arbitrary matter field:
\begin{eqnarray}
& &\frac{1}{\sqrt{-g}}\bigg[-\nabla_\lambda\left(\sqrt{-g}f'(R)g^{\mu\nu}\right)+\nabla_\sigma\left(\sqrt{-g}f'(R)g^{\mu\sigma}\right){\delta^\nu}_\lambda\bigg]+{}\\ & &\qquad\qquad+2f'(R)g^{\mu\sigma}S^{\phantom{ab}\nu}_{\sigma\lambda}=\kappa(\Delta_{\lambda}^{\phantom{a}\mu\nu}-\frac{2}{3}\Delta_{\sigma}^{\phantom{a}\sigma[\nu}{\delta^{\mu]}}_{\lambda})
\end{eqnarray}
The form of this equation implies that even if $f$ is linear in $R$ the connection is not metric compatible for all matter fields because $\Delta_{\lambda}^{\phantom{a}\mu\nu}\neq 0$. To understand how torsion is introduces let us check the simple case where
\be
\label{dsym}
\Delta_{\lambda}^{\phantom{a}[\mu\nu]}=0.
\ee
After some manipulations of the field equations this leads to \cite{sotlib}
\be
S^{\phantom{ab}\nu}_{\sigma\lambda}=0,
\ee
and, therefore, torsion  vanishes if there is no spin current ($\Delta_{\lambda}^{\phantom{a}[\mu\nu]}$). We can safely conclude that specific types of matter field --- those that lead to a non symmetric hypermomentum --- will introduce torsion. Example of such fields are massive vector fields, Dirac field but also an imperfect fluid. We can then conclude that torsion is introduced by particles with a spin  at a microscopic level. The photon is excluded and this remarkably agrees with what we already know from quantum field theory, i.e. that photon are not described by their spin but their helicity.
%%%%%%%%%%%%%%%%%%%%%%%%%%%%%%%%%%%%%%%%%%%%%%%%%%%%%%%
\section{Conclusions}
\label{conc}
%%%%%%%%%%%%%%%%%%%%%%%%%%%%%%%%%%%%%%%%%%%%%%%%%%%%%%%

We have given here, following the spirit of \cite{sotlib} the basics of the metric-affine formalism of $f(R)$ gravity. It is important to note once more that metric-affine gravity is a modification of GR that follows its geometrical spirit but relaxes its simplifying assumptions. This way a much wider phenomenology is achieved. As we saw by considering some examples of the most standard matter fields, metric-affine $f(R)$ gravity has similar physical predictions with GR in the most well studied cases, which is an advantage given the fact that there exist serious experimental bounds in the behaviour of viable gravitational theory. However, at the same time the phenomenology is drastically different in less studied cases, including torsion and non-metricity. Therefore, we believe that metric-affine gravity can pose an interesting alternative to GR, and address, at least partially, some open theoretical and experimental problems related to gravity. Of course, numerous details of the theory are yet to be examined, and it is still to early to know whether metric-affine $f(R)$ gravity will turn out to be a successful gravitational theory. However, we are convinced that studying it further will give  a deeper insight in the nature of the gravitational interaction, which seems to be one of the main puzzles in modern physics.

%%%%%%%%%%%%%%%%%%%%%%%%%%%%%%%%%%%%%%%%%%%%%%%%%%%%%%%%
%\ack The authors would like to thank...

\end{document}